\documentclass[fleqn,usenatbib]{mnras}

\usepackage{newtxtext,newtxmath,bm}
\usepackage[T1]{fontenc}

\DeclareRobustCommand{\VAN}[3]{#2}
\let\VANthebibliography\thebibliography
\def\thebibliography{\DeclareRobustCommand{\VAN}[3]{##3}\VANthebibliography}

\usepackage{graphicx}	% Including figure files
\usepackage{amsmath}	% Advanced maths commands

\providecommand{\mnras}{Mon. Not. R. Astron. Soc.}
\providecommand{\apj}{Astrophys. J.}
\providecommand{\apjl}{Astrophys. J. Lett.}
\providecommand{\apjs}{Astrophys. J. Suppl. Ser.}
\providecommand{\aap}{Astron. Astrophys.}
\providecommand{\pre}{Phys. Rev. E}

\newcommand{\avgZ}{\langle Z\rangle}         % mean charge <Z> = n_e/n
\newcommand{\msD}{\mathcal{D}}               % Maxwell–Stefan pair diffusivity (\msD_{ij})
\newcommand{\Dmat}{[D]}                      % Fick matrix (elements D_{ij}); tracer coefficients are D_i
\newcommand{\aE}{a_{e}}                      % electron-sphere radius
\newcommand{\taud}{\tau_{0}}                 % time unit
\newcommand{\GMCP}{\Gamma_{\rm MCP}}         % mixture coupling parameter

\title[Uphill Diffusion in Yukawa Plasmas]{Uphill Diffusion in Yukawa Plasmas}

\author[M. E. Caplan, E. B. Bauer,  \& M. Bettioua]{
M. E. Caplan,$^{1}$\thanks{E-mail: mecapl1@ilstu.edu}
Evan B. Bauer,$^{2}$
M. Bettioua$^{1}$
\\
$^{1}$Department of Physics, Illinois State University, Normal, IL 67190, USA\\
$^{2}$Lawrence Livermore National Laboratory, Livermore, CA 94550, USA
}

\date{Accepted XXX. Received YYY; in original form ZZZ}

\pubyear{\the\year{}}

\begin{document}
\label{firstpage}
\pagerange{\pageref{firstpage}--\pageref{lastpage}}
\maketitle

\begin{abstract}
In a fluid mixture of many species an individual species can diffuse toward higher concentration without any external driving force because its flux is coupled to the gradients of other components. Such `uphill diffusion' is well known in the chemical and metallurgy literature but to our knowledge has not been studied for strongly coupled plasmas in neutron star and white dwarf interiors. We have previously developed a semi-analytic diffusion model that accurately predicts diffusion coefficients of individual species in a mixture in equilibrium. Uphill diffusion allows us to test this model out of equilibrium, and also test a generalized Darken relation for predicting diffusivities. We derive an analytic theory and compare to three-component molecular dynamics simulations and find very good agreement. This provides a novel out-of-equilibrium test of the generalization of our diffusion model to arbitrary multi-component mixtures. 
\end{abstract}


%%%%%%%%%%%%%%%%%%%%%%%%%%%%%%%%%%%%%%%%%%%%%%%%%%

%%%%%%%%%%%%%%%%% BODY OF PAPER %%%%%%%%%%%%%%%%%%

\section{Introduction}

Fick's first law of diffusion states plainly that a species will diffuse from a region of high concentration to low concentration, finding equilibrium when the concentration gradient is eliminated. The same intuition does not hold in a mixture of more than two components, and a concentration of one species that is initially constant can spontaneously develop a gradient if the other species in the mixture are themselves out of equilibrium with steep concentration gradients. The mechanism for this becomes more clear when recalling that diffusion is best described as being driven by gradients in chemical potential (rather than concentration alone), and that the fluid species exchange momentum with each other through collisional drag terms.  Friction from the background species diffusing therefore creates a diffusive bias, causing the smoothly distributed species to develop a gradient where none existed prior.
These effects have been discussed extensively in the metallurgy and chemical engineering literature \citep{darken1942diffusion,taylor1993multicomponent,Krishna2019}, and have been demonstrated with metal alloys welded together \citep{darken1949diffusion} and mixtures of gasses in bulbs \citep{duncan1962}, among many others. Here we develop a description of diffusion in astrophysical strongly-coupled plasmas and test that it can properly account for inherently multi-component phenomena such as `uphill diffusion' in mixtures of three or more species.

\cite*{CBF}, hereafter CBF, present a model for diffusion in Yukawa plasma mixtures (see Sec. \ref{sec:theory}). Following CBF and \cite{Bildsten2001}, diffusing nuclei should obey the Einstein-Stokes relation because they are spheres with a radius $R$ determined by the nuclear charge diffusing against a background with viscosity $\eta$ such that $\label{eq:Stokes} D = {k_{\rm B} T}/{4 \pi \eta R}$. These self diffusion coefficients are easy to get with MD, but it is not trivial or obvious how to map them onto $D_{ij}$ interdiffusivities for a mixture; in fact, it is not even clear a priori that $\{D_{i}\}$ alone are sufficient to populate the off diagonal elements of the $D_{ij}$ matrix. The CBF model describes diffusion at moderate to strong coupling in terms of Stokes-Einstein drift in an averaged background plasma with one viscosity seen by ions of all charges. In this paper, we show that this insight, coupled with a generalized Darken relation
\begin{equation}
\msD_{ij} = \frac{D_i D_j}{D_{\rm mix}},\; \frac{1}{D_{\rm mix}} \equiv \sum_k \frac{x_k}{D_k}~,
\label{eq:generalDarken}
\end{equation}
provides a closed system of equations that fully describes diffusion in a multi-component, strongly-coupled Yukawa system of three species \citep{White2019}. This is sufficient to describe even non-trivial diffusion behavior such as ``uphill'' diffusion of a species against its own concentration gradient due to coupling of its flux to the other species, i.e. an interspecies friction that is encoded in $\msD_{ij}$.

We are not so concerned with contriving a specific astrophysical scenario in this work, but such situations can arise in nature. For example, buoyant crystals formed by distillation melt into a liquid of different composition \citep{Blouin2021,Bedard2024}; compositional convection in a neutron star ocean can create interfaces between different liquid compositions \citep{Medin2011}; a partly remelted accreted crust is in contact with the liquid that deposited it \citep{Horowitz2007,Mckinven2016}, and so on. Most generally, the evolution of all these mixtures and phenomena have some sort of diffusive component. Reliable diffusion coefficients are therefore necessary for modeling this out of equilibrium behavior. However, diffusion coefficients themselves are most reliably determined from equilibrium molecular dynamics simulations. Uphill diffusion is therefore interesting to study, both in its own right as a novel phenomena, and also as a practical test of the existing diffusion models applied to out of equilibrium fluids of broad astrophysical importance.

\section{Analytic Theory}\label{sec:theory}

\subsection{Single-Species Diffusion in a Uniform Background}
\label{sec:CBF}

For a single species in one dimension $z$, Fick's laws provide the most intuitive understanding of diffusion driven by out-of-equilibrium composition gradients. When diffusing in a uniform background in the absence of external forces with positive coefficient $D_i$, Fick's first law states that the flux is $J_i = -nD_i dx_i/dz$, where $x_i =n_i/n$ is the number fraction of species $i$. That is, in this one-component case, the flux always points `downhill' in the concentration, with its overall rate set by a single positive coefficient $D_i$ that encodes the physics of how that species exchanges momentum with its surroundings. 

Our previous work, \citet*{CBF} (hereafter CBF) describes a semi-analytic model for constructing the single-species (tracer) diffusion coefficients $D_i$ in an uniform strongly-coupled mixture. That work fit one component plasma (OCP) MD with a function $D^*_{\rm OCP}(\GMCP,\kappa)$, $\GMCP=\langle Z^{5/3}\rangle e^2/(\aE k_BT)$, and extended it to mixtures by treating each ion as a Brownian particle in a fluid with the viscosity of the equivalent one-component plasma,
\begin{equation}
D_i=D_0\,Z_i^{-0.6},\qquad D_0\equiv D^*_{\rm OCP}(\GMCP,\kappa)\,\langle Z \rangle^{0.6}\,\omega_pa^2,
\label{eq:cbf}
\end{equation}
with $\omega_p$ the plasma frequency. The law reproduces the MD coefficients to 10--15\% for $10\lesssim\GMCP\lesssim200$ \citep{caplan2021mnras,CBF,Caplan2023RNAAS}, with best performance for low $Z_i/\langle Z\rangle$.
The prefactor $D_0$ encodes the physics of viscosity experienced by all particle species and will set overall timescales, while the CBF model postulates that the interaction of each species with that viscosity is set by an effective charge-sphere radius that depends on $Z_i$ alone.

\subsection{Multi-Species Diffusion in a Dynamically Evolving Mixture}
\label{sec:multiTheory}

For a non-uniform background in a multicomponent mixture of $N$ species, it is possible to construct a generalized form of Fick's first law \citep{taylor1993multicomponent}:
\begin{equation}
    (J) = - n[D] (\nabla x)~,
    \label{eq:FickFlux}
\end{equation}
where now $(J)$ and $(\nabla x)$ are vectors of $N-1$ fluxes and number fraction gradients, and $[D]$ is an $(N-1)\times(N-1)$ matrix of diffusion coefficients (the $N$th entry in the system of equations is eliminated by mass conservation). We call $[D]$ the `Fick matrix' and describe its entries $D_{ij}$ as the `Fick coefficients.' As we shall show, it is non-trivial to construct the Fick matrix even when all single-species coefficients $D_i$ are known for the uniform case, and the off-diagonal entries of $[D]$ mean that the composition gradients of {\em all species} can influence the flux of each individual species. This is the origin of `uphill diffusion' in the Fick picture, where the off-diagonal terms in $[D]$ mean that $J_i$ is no longer constrained to have a sign corresponding to the `downhill' direction set by $\nabla x_i$ alone, even in the absence of external driving terms (such as the gravitational forces and ionic coupling chemical potential terms that drive mass stratification in WDs and neutron stars).

For a multi-species plasma mixture, it is often more natural to express the coupled set of diffusion equations in a form like the Maxwell-Stefan equations \citep{Maxwell1867,Stefan1871}:
\begin{equation}
-\nabla\tilde\mu_i=\sum_{j\neq i}\frac{x_j\,(\bm v_i-\bm v_j)}{\msD_{ij}},\qquad \tilde\mu_i\equiv\frac{\mu_i}{k_BT},
\label{eq:MS}
\end{equation}
where $\mu_i$ are the species chemical potentials, $\bm v_i$ are the species velocities, and we label $\msD_{ij}$ the `Maxwell-Stefan' coefficients. This form corresponds more closely to the diffusion equations that arise from kinetic theory approaches such as the Chapman-Enskog method \citep{Chapman1970} and the Burgers equations \citep{Burgers1969} which are commonly used for stellar diffusion calculations. Here we have simplified relative to those approaches by dropping the thermal diffusion terms, which are not relevant for strong degeneracy and coupling. The Maxwell-Stefan coefficients map directly onto the Burgers `resistance coefficients' as $K_{ij} = n_in_jk_BT/n\msD_{ij}$. In Eq.~\eqref{eq:MS} the influence of each species on the others can be thought of an inter-species friction that adds linearly to form the total frictional force in the sum on the right hand side. In principle the $\mu_i$ in Eq.~\eqref{eq:MS} could contain driving terms such as gravity, electric fields, and charge-dependent chemical potential corrections that drive sedimentation in compact stars, but we will deliberately neglect these  in both our calculations and the MD so that the long-term equilibrium is always a uniformly mixed system by construction here. This allows us to isolate the impact of the interaction terms in the dynamically evolving mixture most relevant for understanding the full mixture $\msD_{ij}$.

It is not obvious {\em a priori} how to translate from the single-species equilibrium coefficients $D_i$ of Section~\ref{sec:CBF} to either of the more general multi-species formulations with off-diagonal inter-diffusion coefficients: $D_{ij}$ for Eq.~\eqref{eq:FickFlux} or $\msD_{ij}$ for Eq.~\eqref{eq:MS}. Here we propose that the most natural way to formulate mixture coefficients within the CBF model is the generalized Darken relation for the Maxwell-Stefan coefficients given by Eq.~\eqref{eq:generalDarken}. We note that Eq.~\eqref{eq:generalDarken} has the desirable property that $\sum_j x_j/\msD_{ij} = 1/D_i$, so that for a trace species diffusing in a uniform mixture of many other species ($\bm v_j \approx 0$), Eq.~\eqref{eq:MS} reduces to $\bm v_i = -D_i\nabla\tilde\mu_i$. A trace species therefore diffuses according to its tracer diffusion coefficient $D_i$ as desired in this limit, while the full $\msD_{ij}$ can describe multi-component diffusion far from equilibrium as we will explore in later sections.
Although we are not aware of the generalized Darken relation of Eq.~\eqref{eq:generalDarken} appearing broadly under that name in the literature, a few references do explore its validity. 
Eq.~\eqref{eq:generalDarken} for the Maxwell-Stefan coefficients appears in \cite{liu2011}, where they derive it for $N=2-4$ by neglecting velocity cross-correlations between distinct particles, which naturally matches the physical picture presented by CBF. 
\cite{White2019} explore the applicability of Eq.~\eqref{eq:generalDarken} with MD simulations in the warm dense matter regime, and generally find good agreement for ternary mixtures, though the agreement degrades somewhat for components with highly asymmetric masses in their study.
Eq.~\eqref{eq:generalDarken} can also be shown to be equivalent to a multi-component Darken relation for the Fick matrix developed much earlier in the metallurgical literature (e.g., \citealt{AnderssonAgren1992}).
In the case of precisely two species, Eq.~\eqref{eq:generalDarken} reduces to the original relation of \cite{darken1948diffusion}: $\msD_{12} = x_2 D_1 + x_1 D_2$. For two species and an ideal chemical potential, it can be shown that Eq.~\eqref{eq:MS} readily reduces to $J_1 = -n\msD_{12}dx_1/dz$, so that one coefficient $\msD_{12}$ describes the whole system and still produces a result that matches the intuition of Fick's laws. For 3 or more species, inverting Eq.~\eqref{eq:MS} to form Eq.~\eqref{eq:FickFlux} introduces a matrix with off-diagonal terms, even for ideal potentials.

The relation of Eq.~\eqref{eq:generalDarken} says that in the Maxwell-Stefan picture, the pairwise interaction coefficients depend only on the two tracer coefficients of the interacting species, scaled by an overall number-weighted average of tracer drags in the local background (the inverse sum over $D_k$ averages over drag as opposed to mobility). Under the CBF model in strongly-coupled plasmas, this states more specifically that the pairwise interaction coefficients depend only on their two charge-sphere radii relative to the overall average of the background particles and the mixture viscosity. We will test diffusion results using the coefficients of Eq.~\eqref{eq:generalDarken} against MD simulations that are independent of any assumed coefficient form.

\subsection{Yukawa Plasma in White Dwarfs and Neutron Stars}

The interiors of young white dwarfs and the oceans of neutron stars are liquids of fully ionized nuclei immersed in a degenerate electron gas interacting through the screened Coulomb potential ${V_{ij}=Z_iZ_je^2e^{-r/\lambda}/r}$ with nuclear charges $Z_i$ and $Z_j$ separated by distance $r$ with screening length $\lambda$. The liquid is strongly coupled, with $\Gamma=Z^2e^2/(ak_BT)$ between 1 and the crystallization value $\approx175$, with $a$ the ion-sphere radius and screening $\kappa=a/\lambda\approx0.1-1$. We report lengths in the electron-sphere radius $\aE=(3/4\pi n_e)^{1/3}$ and times in $\taud=(m_u\aE^3/e^2)^{1/2}$. 

For this work we consider a Yukawa plasma at uniform temperature $T$ with species number densities $n_i$ and charge density $n_e=\sum_iZ_in_i$, such that the charge fractions are $Y_i=Z_in_i/n_e$ with $\sum_iY_i=1$ and mean charge $\langle Z \rangle=\sum_i x_i Z_i$.

Multiplying Eq.~\eqref{eq:MS} by $x_i$, writing the velocities as charge fluxes $q_i=Z_in_iv_i$ and using $x_i=Y_i\avgZ/Z_i$ puts Eq.~(\ref{eq:MS}) in the form we use, effectively a variable change of $x\to Y$ and $J\to q$,
\begin{equation}
-\frac{Y_i}{Z_i}\nabla\tilde\mu_i=\frac1{n_e}\sum_{j\neq i}\frac{Y_jq_i-Y_iq_j}{\msD'_{ij}},\qquad
\msD'_{ij}\equiv\frac{Z_iZ_j}{\avgZ}\,\msD_{ij},
\label{eq:MSq}
\end{equation}
with $\sum_iq_i=0$ because $n_e$ is uniform (no net charge current).

We now obtain $\nabla \tilde\mu_i$ for the Yukawa plasma from the well known free energy, following \cite{Medin2010}. In the linear mixing approximation each species contributes to the free energy density as in a one component Yukawa plasma at the same electron density,
\begin{equation}
\frac FV=k_BT\sum_kn_k\Big[\ln(n_k\Lambda_k^3)-1+f_k\Big]+F_e(n_e),
\label{eq:F}
\end{equation}
with $\Lambda_k$ the thermal wavelength, $f_k\equiv f(\Gamma_k,\kappa_k)$ the excess free energy per ion, in units of $k_BT$, of the Yukawa plasma of species $k$ at coupling $\Gamma_k=Z_k^{5/3}e^2/\aE k_BT$ and screening $\kappa_k=Z_k^{1/3}\aE/\lambda$, and $F_e$ the free energy of the electron background. 
At fixed $T$, $\Gamma_k$ and $\kappa_k$ depend on the composition only through $n_e$.

For our toy problem here, and when comparing to our Yukawa MD simulations, we treat the
degenerate electron background as effectively incompressible, so that $n_e$ is uniform.
The ion densities are then not independent: charge neutrality requires
$\sum_i Z_i n_i = n_e$ at every point. Taking chemical potentials subject to this
constraint introduces a Lagrange multiplier $\chi$, and because the constraint involves
each species only through its charge $Z_i n_i$, the multiplier enters every chemical
potential in the same way,
\begin{equation}
  \mu_i = k_B T \left[ \ln(n_i \Lambda_i^3) + f_i \right] + Z_i \chi .
  \label{eq:mu}
\end{equation}
Physically, $Z_i\chi$ is the response of the electron background, acting on each ion in
proportion to its charge. Its value resembles the electron chemical potential, but its
gradient is set by mechanical equilibrium rather than the electron equation of state,
as determined below.

Because $\Gamma_i$ and $\kappa_i$ depend on composition only through $n_e$, the non-ideal term $f_i$ then drops out of $\nabla\mu_i$, leaving
\begin{equation}
  \nabla\tilde\mu_i = \nabla\ln n_i + Z_i\nabla\tilde\chi ,
  \qquad \tilde\chi \equiv \frac{\chi}{k_B T} .
\end{equation}
The $\nabla\tilde\chi$ term can be determined by summing Eq.~\eqref{eq:MS} over species: multiplying by $x_i$ and summing, the right-hand side vanishes by antisymmetry in $i \leftrightarrow j$, so
\begin{equation}
  0 = \sum_i x_i\nabla\tilde\mu_i
    = \sum_i x_i\nabla\ln n_i
      + \Big(\sum_i x_i Z_i\Big)\nabla\tilde\chi .
\end{equation}
Since $x_i = n_i/n$, the first sum is $n^{-1}\sum_i \nabla n_i$, and the second is $\sum_i x_i Z_i = n_e/n$, so $\nabla\tilde\chi = -\sum_i \nabla n_i/n_e$.

It will be convenient to express everything in terms of dimensionless charge fractions, so using $n_i = Y_i n_e/Z_i$ gives $\nabla\ln n_i = \nabla Y_i/Y_i$ and
\begin{equation}
\nabla\tilde\mu_i=\frac{\nabla Y_i}{Y_i}-Z_i\sum_k\frac{\nabla Y_k}{Z_k}.
\label{eq:gradmu}
\end{equation}
This can be understood simply as an entropic force for mixing plus an additional pressure that pushes ions, more strongly with higher charge, toward the side with fewer ions per unit charge. This has the added benefit of being entirely in terms of dimensionless quantities. Qualitatively, Eq.~\eqref{eq:gradmu} already suggests conditions to simulate: the uphill diffusion will be strongest for greatest variations in $\nabla\tilde\mu_i$ which can be obtained from large charge ratios and large steps in $Y_i$ across the interface.

\subsection{Uphill Example: Three-Species Diffusion}
\label{sec:analyticUphill}

Consider three species of nuclei with charges $Z_1,Z_2,Z_3$. Throughout this work species 1 and 2 occupy $z<0$ and species 2 and 3 occupy $z>0$. Species 2, the `uphill' species, has the same charge fraction $Y_2$ on both sides ($Y_2^{z<0} = Y_2^{z>0}$). The previous section provides all the machinery needed to solve for the time-dependent composition evolution numerically: Eqs.~\eqref{eq:MSq} \& \eqref{eq:gradmu} just need an expression for the pair diffusivities in the Yukawa plasma, which are supplied by Eqs.~(\ref{eq:generalDarken}--\ref{eq:cbf}), the generalized Darken relation applied to the CBF coefficient model.

We now proceed to map out an approximate analytic solution by transforming the coefficients described in Section~\ref{sec:multiTheory} to the applicable Fick matrix.
Transforming our Darken relation into the Fick matrix form yields more complex expressions for the coefficients than the Maxwell-Stefan formulation in Eq.~\eqref{eq:generalDarken}, but has the advantage of expressing flux and composition evolution explicitly as in Eq.~\eqref{eq:FickFlux} rather than only implicitly in a coupled system of equations as in Eq.~\eqref{eq:MS}. This allows introducing approximations that yield closed-form analytic solutions for quantities of interest that will motivate exploring particular mixture configurations, and we can compare the results using our approximations to both numerical solutions of the fully-coupled system in Eq.~\eqref{eq:MS} and MD simulations.
We obtain the Fick matrix coefficients by recalling the $\sum_i q_i =0$ because we expect charge conservation so no net charge flux. Using ${Y_3 = 1-Y_1-Y_2}$ to eliminate variables, and working in the frame of no net charge flux ($x\to Y, J\to q$), we obtain the matrix diffusion equation $\partial_t \bm Y=\partial_z(\Dmat\partial_z\bm Y)$ for the two independent charge fractions with matrix elements
\begin{equation}
\;\begin{aligned}
\frac{D_{ij}}{D_0}={}&Z_i^{-0.6}\delta_{ij}-Y_i\big(Z_j^{-0.6}-Z_3^{-0.6}\big)\\
&-Y_i\Big(Z_i^{0.4}-\sum_kY_kZ_k^{0.4}\Big)\Big(\frac1{Z_j}-\frac1{Z_3}\Big)
\end{aligned}\; 
\qquad i,j\in\{1,2\}.
\label{eq:Dij}
\end{equation}

\noindent Following from the structure of the CBF law, $\Dmat/D_0$ is dependent solely on the two dimensionless quantities $Y_i$ and $Z_i$ while $D_0$ sets a scale, and eigenvalues $\lambda_{1,2}=\bar\lambda\pm\sqrt{\tfrac14\,(D_{11}-D_{22})^{2}+D_{12}D_{21}}$ with mean $\bar\lambda=\tfrac12\,(D_{11}+D_{22}).$ Darken does make it possible to obtain analytic expressions for the matrix for arbitrary $N$, see Appendix \ref{sec:appendixa}.

The off diagonal element $D_{21}$ is responsible for uphill diffusion, and the magnitude of the overshoot can be found analytically. Consider an interface in an infinite medium of two binary mixtures, with $(Y_1, Y_2,0)$ at $z<0$ and $(0, Y_2, Y_3)$ at $z>0$. If we fix $\Dmat$ to the mean composition $\bar Y_i = \tfrac12 (Y_i^{z>0} + Y_i^{z<0}) $ the solution is an error function of $\Dmat$; note that this is equivalent to a fully mixed state, hereafter $[D(t=\infty)]$. With the one off diagonal element $D_{21}$ the uphill pulse shape is simply

\begin{equation}
    \delta Y_2 (z,t) = -\frac{1}{2}\frac{D_{21}\Delta Y_1}{\lambda_1-\lambda_2}
\left[
\operatorname{erf}\left(\frac{z}{2\sqrt{\lambda_1 t}}\right)
-
\operatorname{erf}\left(\frac{z}{2\sqrt{\lambda_2 t}}\right)
\right]
\end{equation}
\noindent and matches the dashed curve obtained numerically in Fig.~\ref{fig:infmedium}. For $N > 3$ there is a slightly more complicated analytic solution but it is similarly a sum of error functions of eigenvalues, see Appendix \ref{sec:appendixa}.

The uphill species develops an excess dependent only on $D_{21}$ because $\Delta Y_2 =0$ across the interface. Species 2 develops a peak excess $\delta n_2 / n_2$ on one side and an equal deficit on the other of fixed amplitude that spreads diffusively like $\sqrt t$ such that we define

\begin{equation}
\;\frac{\delta n_2}{n_2}\Big|_{\max}=0.121\,\Pi\ \ \text{at}\ \ z=\pm\sqrt{2\bar\lambda t},\qquad
\Pi\equiv\frac{|D_{21}\,\Delta Y_1|}{Y_2\,\bar\lambda}\;
\label{eq:Pi}
\end{equation}
\noindent where $0.121=1/(2\sqrt{2 \pi e})$.\footnote{The prefactor 0.121 uses the approximation that $\lambda_1=\lambda_2.$ For the fiducial O/Fe/Se mixture in the next section the actual value is $0.125 \Pi$.} Observe that $\Pi$ is dimensionless such that the magnitude of the overshoot depends solely on $Z_i$ and $Y_i$, independent of temperature, density, and screening. The sign $D_{21}\Delta Y_1>0$ drives an excess to $z<0$ because we choose $Z_1 < Z_3$; the uphill excess is observed to move toward the lower charge species.

\section{Numerical Simulations}\label{sec:numericalsims}

We now perform simple numerical simulations and compare with molecular dynamics. Mixtures that maximize the overshoot $\Pi$ can be intuited from inspection of $D_{ij}$ in Eq.~\eqref{eq:Dij}; first, choose $Y_i \sim 0.5$ to produce large initial concentration gradients while still giving good statistics in real molecular dynamics, then, the parenthetical terms immediately suggests large differences in $Z_i$ are preferred.
We consider the specific case of O-Se in contact with Fe-Se so that Se flows into the O. In our notation this is $Z = (8, 34, 26)$. We use $Y^{z<0} = (0.7, 0.3, 0)$ and $Y^{z>0} = (0, 0.3, 0.7).$ The temperatures are the same but the plasma coupling parameter $\Gamma_{\rm MCP}$ varies locally with composition so the true diffusion coefficients vary across the interface and time.

\begin{figure}
 \includegraphics[width=0.9\columnwidth]{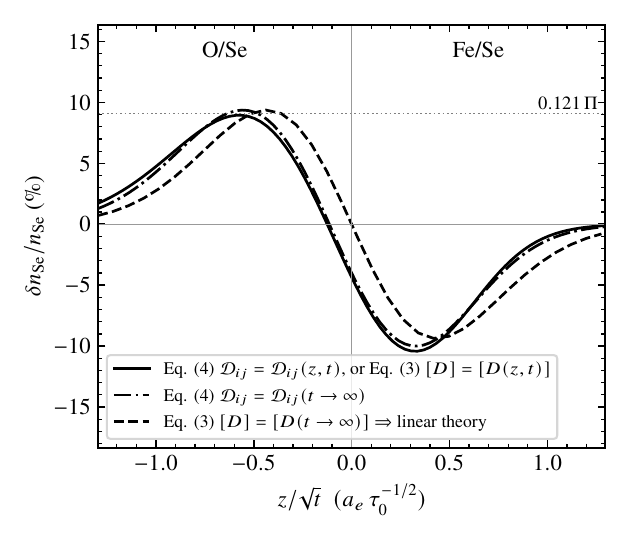}
 \caption{A numerical simulation of the interface between an O/Se and Fe/Se volume, initially with constant Se concentration $Y_{2}=0.3$ on both sides, spontaneously develops an excess of Se. This pulse has a constant excess that shifts leftward and broadens in time, but is scale free in diffusive time $\sqrt{t}$. The expected overshoot is roughly 10\% for both a self consistently calculated ${[D]}$ based on the local composition (${[D(z,t)]}$) and for the linear theory with a constant composition taken to be the mixture average ($[D(t\to\infty)]$). The only notable qualitative difference is a slight loss of symmetry where the zero of the self-consistent calculation is shifted left into the O side. The simulation is at constant temperature with initial $\Gamma_{\rm MCP}^{\rm O/Se}=35.7$ and $\Gamma_{\rm MCP}^{\rm Fe/Se} = 150$ at $\kappa=0.46.$ The curves are nearly identical when slightly warmer at $\Gamma_{\rm MCP}^{\rm Fe/Se}=100$.}
 \label{fig:infmedium}
\end{figure}

\begin{figure*}
 \includegraphics[width=\textwidth]{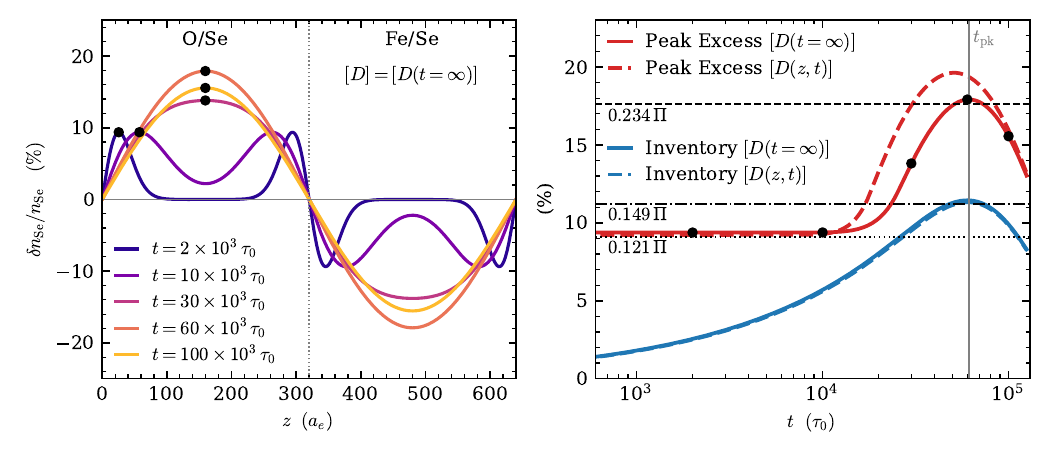}
 \caption{(Left) Numerical simulation of the Fick matrix in a periodic box. The pulse of excess Se from each interface moves inward and broadens  at early times as in the case of the infinite medium of Fig.~\ref{fig:infmedium}, but when the pulses meet at intermediate times they constructively interfere to form something like a sine wave before decaying away as the mixture homogenizes. Black points show the peak excess at each time. (Right) The peak overshoot (red) at early times is basically the same for the self consistently tracked $[D(z,t)]$ compared to the mean mixture $[D(t=\infty)]$, but because the self consistent $\Dmat$ pulse moves inward more quickly it is able to reach a slightly greater peak at intermediate times. However, the excess of species 2 in the left half of the box which we call the `inventory' (blue) is largely unaffected by our choice of $[D]$, as discussed in the text. 
 Animations available in the supplemental materials.}
 \label{fig:periodicsimple}
\end{figure*}

Figure \ref{fig:infmedium} shows that the overshoot predicted by linear theory and the overshoot we should expect in a true molecular dynamics simulation are not so different, and qualitatively the same but weakly shifted. The figure shows the early-time evolution at the interface between the two regions for a self-similar spatial coordinate $z/\sqrt{t}$, considering three cases for numerically solving the three-component system:
\begin{enumerate}
\item A self-consistent numerical solution of the Maxwell-Stefan equations in Eq.~\eqref{eq:MS} that recomputes the local $\msD_{ij}$ for each slice in $z$. This is also the same as solving the equivalent Fick formulation of Sec.~\ref{sec:analyticUphill} with $[D(z,t)]$ dynamically reflecting the evolving composition.
\item A numerical solution of Eq.~\eqref{eq:MS} using coefficients for the mean mixture $\msD_{ij}(t\to\infty)$ as an approximation.
\item A solution of the Fick formulation using the mean mixture $[D(t\to\infty)]$, which matches the derivation of $\Pi$ in Sec~\ref{sec:analyticUphill}.
\end{enumerate}
The first of these is the highest physical fidelity and will be useful for comparing to the MD. The last is the analytic approximation most suitable for comparing to the equations of Sec.~\ref{sec:analyticUphill}.
All curves reach approximately the same peak overshoot but with the fully self-consistent $\Dmat$ resulting in slightly faster diffusion, but the lobes are still self similar in $\sqrt{t}$. The intercept, $\delta n_2 /n_2 = 0$ is slightly displaced to the left. We conclude that an overshoot of about 10\% may be likely in this case, consistent with our prediction of $\Pi$ from linear theory, and can be resolved with an MD simulation of about $10^5$ particles.\footnote{Consider a box subdivided into 100 `slices' in $z$ to obtain histogram counts of the uphill species. With order a few $\times10^4$ particles of $Z_2$ there may be a few $\times10^2$ in each `slice' with Poisson noise of order $10^1$, so with averaging and smoothing an uphill excess of order $0.121\Pi \sim 10^{-1}$ can be resolved from MD configurations sampled closely in time.}

To understand why the formulation of (ii) more closely matches (i) than (iii) does,
we note that the Maxwell--Stefan formulation places kinetics and thermodynamics in
separate terms: the $\mathcal{D}_{ij}$ describe interspecies friction, the
composition enters as explicit weights (the friction on species $i$ is a sum
over partners $j$, each weighted by its number fraction $x_j$), and the free
energy enters only through the driving forces $\nabla\tilde{\mu}_i$. 
In the CBF + Darken model the coefficients take the form
\begin{equation}
  \frac{\mathcal{D}_{ij}}{\omega_p a^2}
  = D^*_{\rm OCP}(\Gamma_{\rm MCP},\kappa)
    \left(\frac{Z_i Z_j}{\langle Z\rangle^2}\right)^{-0.6}
    \left\langle\left(\frac{Z}{\langle Z\rangle}\right)^{0.6}\right\rangle ,
  \label{eq:Dij_CBF_Darken}
\end{equation}
so once the background state ($\Gamma_{\rm MCP}$, $\kappa$, $\langle Z\rangle$)
is specified, each coefficient depends only on the charges of the two species
relative to the mean. 
The Fick matrix, by contrast, combines friction,
thermodynamics, and the individual charge fractions into a single object,
$[D] = [B]^{-1}[G]$ in the notation of \citet{taylor1993multicomponent}, where $[B]$ is the
friction matrix and $[G]$ the thermodynamic factor. This formulation carries
the local composition weights through into $\Dmat$ itself, even for an ideal mixture.
Composition weight terms clearly appear in the expression for the $\Dmat$ coefficients
in Eq.~\eqref{eq:Dij}. 
Consistent with this picture,  holding the $\mathcal{D}_{ij}$ fixed at
their mean-mixture values while evolving the composition weights reproduces
the self-consistent solution closely, whereas the linear theory, with $[D]$
frozen at the mean mixture, does not match as closely but still produces
a comparable peak and qualitatively similar behavior.

In Figure \ref{fig:periodicsimple} we show this pulse for the same O/Se \& Fe/Se mixtures in a periodic box, obtained by numerically integrating the diffusion equation on a fine grid. We show a full evolution of the pulse for just the linear theory with the simplified $\Dmat$; a self consistent simulation is qualitatively the same but with times slightly shifted. An animation comparing them is available in the supplemental materials. For a box of this length, approximately $z=630 a_e$, we expect the two pulses to move inward into the O-rich side and broaden with equal peak excesses of about $0.12 \Pi$ until they meet after about $10^4 \, \tau_0$. The peaks then merge and the peak overshoot grows up to about $0.24 \Pi$ until they resemble something like the fundamental mode of the Fourier series before decaying away as the box homogenizes.  

We define the `inventory' as the the excess of species 2 in the left half of the box and show how it evolves for two cases on $\Dmat$: (i) a fixed composition using the mean mixture so that $[D(t\to\infty)]$ is constant as in the derivation of $\Pi$ above and (ii) a self-consistent simulation that recomputes the local $D_{ij}$ for each slice in $z$ so that $[D(z,t)]$ is dynamic, which should better reproduce the MD. Despite the differences in peak excesses the inventories are largely unchanged and this can be understood simply. The inventory is equivalent to the integrated flux across the interface, and the interface immediately adopts the mixture average $\bm Y = (0.35, 0.3, 0.35)$. The fluxes should therefore be very similar. There is a small difference between $[D(z,t)]$ and $[D(t=\infty)]$ because of the left-shift seen in the self consistent simulation, but this is a relatively small change in composition. Furthermore, any excess inventory caused by the difference in composition and flux at the interface is then partially canceled by the negative contribution of $\delta n_{\rm Se}/n_{\rm Se}$ from the left shifting (see center, Fig.~\ref{fig:infmedium}). While one could plausibly calculate the inventory considering only the region between the zero crossings of $\delta n_{\rm Se}/n_{\rm Se}$, the noise in a real molecular dynamics simulation will make this challenging and so we prefer to use the positions of the initial interfaces as our boundaries (\textit{i.e.} the Matano plane). 
%partially offsetting the contribution from the higher peak.

A full molecular dynamics simulation can test this nearly exactly. We consider a mixture with $Z_1 < Z_3$ again so that uphill flow is to the left, with an $(Y_1,Y_2,0)$ fluid spanning $0<z< L/2$ and a $(0,Y_2,Y_3)$ fluid spanning $L/2<z<L$. This construction results in two interfaces, one at $L/2$ and one at the periodic boundaries, such that in a time dependent simulation two pulses form at early times until merging when their peaks approach $L/4$ and $3L/4$ and merge into something like a sine wave in $Y_2$ before the overshoot finally decays away as box homogenizes. Solving the diffusion equation for the periodic condition and taking the fundamental Fourier mode of the initial `square wave' composition one finds a peak overshoot  
\begin{equation}
\frac{\delta n_2}{n_2}\Big|_{\max}\simeq0.234\,\Pi\ \ \text{at}\ \ t_{\rm pk}=\Big(\frac L{2\pi}\Big)^2\frac{\ln(\lambda_1/\lambda_2)}{\lambda_1-\lambda_2}
\label{eq:box}
\end{equation}
with $\lambda_{1,2}$ the eigenvalues of $\Dmat$. This is almost exactly twice what was found for the single interface above ($0.121 \Pi$) because it is just two pulses meeting constructively.\footnote{Discarding the assumption that $\lambda_1 = \lambda_2$ only tends to shift the prefactor of $\Pi$ up by a small amount. For the fiducial O/Fe/Se mixture, it is $0.239 \Pi$.}

The periodic box now gives us another test to consider in the MD. The inventory (excess of species 2 in the left half of the box) also peaks at the same time at a value of $0.149 \Pi$. Because it includes all particles in the left half, it is less sensitive to the small numbers of particles in any small slab in $\Delta z$ used to check the profile. This quantity therefore becomes our strongest test of model.

\begin{figure*}
 \includegraphics[width=0.99\columnwidth]{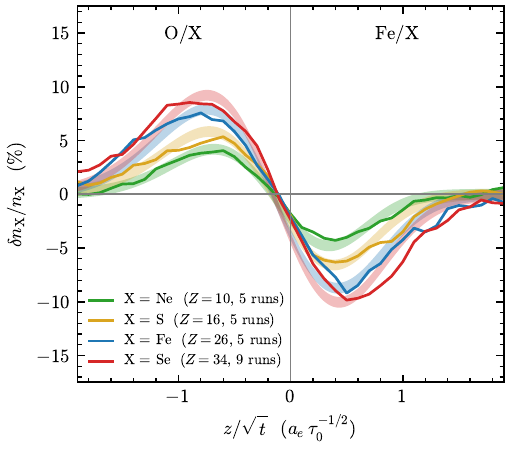}
  \includegraphics[width=0.99\columnwidth]{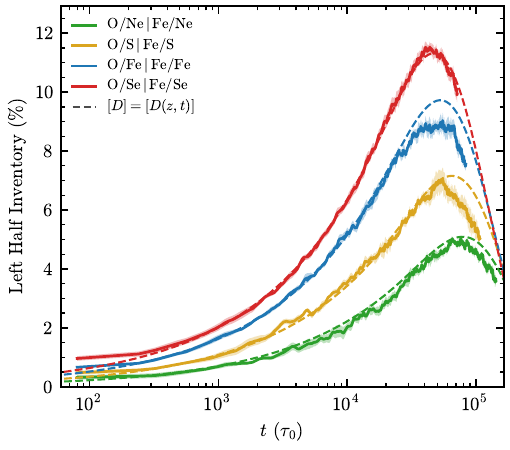}
 \caption{Molecular dynamics simulations of periodic boxes. (Left) At early times before the uphill pulses meet they match the simple infinite medium solution; these curves are the average of 5 independent seed (9 seeds for Se) and combine the pulses from both interfaces from time 0 to 8000 $\tau_0$. Shaded bands show analytic curves for $[D(z,t)]$ with 1\% width. 
 Note that the peak excess is biased slightly upwards by stochastic fluctuations when binning the uphill species, and biased downward by the smoothing, so this is a less exact test than the inventory. (Right) The inventory for our four different mixtures likewise show strong agreement with the linear theory ($[D(t=\infty)]$). An animation for each uphill species is in the supplemental materials.}
 \label{fig:mdinventory}
\end{figure*}

The results of our molecular dynamics simulations with the periodic box are shown in Fig.~\ref{fig:mdinventory}. We consider four uphill species in O/X/Fe mixtures so that our full ensemble tests X = Ne, S, Fe and Se ($Z_X=10,16,26,34$). Each is simulated with five independent seeds for $7.5-9.9\times10^4$ ions per run, except for Se which was simulated with nine seeds to allow us to check for dependence on timestep and to get better statistics for the fiducial composition. We keep the particle counts of the O and Fe backgrounds roughly constant, so for fixed $Y_2$ we have greater numbers of the lower $Z$ species. For the X=Fe case specifically we simply define species 2 and 3 to use identical masses and charges, but they are still tracked as separate populations.

We show the single interface pulses on the left of Fig. \ref{fig:mdinventory}, using style following Fig. \ref{fig:infmedium}. These curves are obtained by averaging over configurations sampled from all our simulations up to time $8000 \tau_0$, before the pulses meet in the center of the O-half of the box, and are rescaled by $\sqrt t$ so we can average them to obtain the the dimensionless pulse. 
Our MD results fall somewhere between the analytic curves for the self consistent calculation $[D(z,t)]$ and the fixed $[D(t\to\infty)]$ and agree with both within noise (analytic curves omitted for readability). The predicted uphill number increases with the charge of the uphill species, with $\Pi=0.34$ for Ne ($0.75$ for Se), giving peak excesses $\delta n_{\rm X} / n_{\rm X}$ for the early pulse of $4.1\%$ ($9.1\%$), matching our simulations within error. At fixed charge fraction $Y_2=0.3$ the total number of particles of the uphill species scales like $1/Z_X$ ($5.0\times10^{3}$ per half box for Se and $1.7\times10^{4}$ for Ne). The weaker signal of the lighter species is therefore offset by better counting statistics.

The inventory for the full simulations are shown on the right in Fig. \ref{fig:mdinventory}. At later times the pulses meet constructively, reach maximum inventory, and decay away. Our simulations were run long enough to resolve peak inventory and some of the post-peak decay. 
The analytic approximation does a terrific job of reproducing the inventories, especially for our O/Fe/Se simulations which has the highest resolution due to the extra simulations. As in Fig.~\ref{fig:periodicsimple}, the inventory is largely insensitive to the choice of analytic model, as the self consistent $[D(z,t)]$ is basically identical to the fixed $[D(t=\infty)]$. 

The analytic inventory curves have an additional empirical correction to the CBF diffusion coefficients. The CBF law is known to underpredict by a few percent in the range of $\Gamma_{\rm MCP}$ we simulate, with the underprediction worsening at higher charge ratios; for example trace Fe in C/O and O/Ne is faster by about 25\% than CBF predicts \citep{Caplan2023RNAAS}. We therefore apply a correction by scaling $D_0$ up for all three species by $1.03$, $1.07$, $1.18$ and $1.29$ for X=Ne, S, Fe, and Se respectively. These values were obtained from the ratio of the $D_i$ of the O in the initial left side box (O/X side) during equilibration compared to the CBF prediction. This causes the peak to shift slightly in time but not does not affect peak inventory because the uphill number $\Pi$ only contains ratios of $D_{ij}$.

\section{Discussion}\label{sec:dis}

We have successfully derived an analytic theory of uphill diffusion for Yukawa plasma mixtures and verified it directly with molecular dynamics simulations. This was obtained combining the Maxwell-Stefan equation with the free energy of the Yukawa fluid, \cite*{CBF} diffusion model, and Darken relation to obtain a Fick matrix entirely in terms of the dimensionless charges and charge fractions. For three species, linear theory matches molecular dynamics, with the inventory (integrated flux) in particular showing fantastic agreement with the theory. 

The test of the generalized Darken relation is also novel, and we know of no other work that provides a test of this relation for mapping from Maxwell-Stefan to Fickian interdiffusivities for Yukawa plasmas.
We have therefore provided the first independent verification of the generalized Darken relation for Yukawa plasmas. 
This success justifies the implementation of the CBF law in codes such as \textsc{MESA} and provides confidence that the implementation that generates the resistance coefficients from the self diffusion coefficients is a reliable one. This also motivates improvements to the CBF model to reduce some of the known systematics - these will be presented in future work.  

Even despite some known weaknesses of the existing CBF law for large charge ratios, our calculations of the uphill species inventory are largely insensitive to this systematic. There is likely a simple explanation; the mixtures only probe a given composition for a brief time, and as a result the total signal is an average over CBF at many local values of $\Gamma$ and $\mathbf{Y}$, causing much of the error to average out.

 \section*{Acknowledgements}

This work was supported by a grant from the Simons Foundation (MP-SCMPS-00001470) to MC. 
Financial support for this publication comes from Cottrell Scholar Award \#CS-CSA-2023-139 sponsored by Research Corporation for Science Advancement.  
Work by EB was performed under the auspices of the U.S.\ Department of Energy by Lawrence Livermore National Laboratory under Contract DE-AC52-07NA27344. Large language models including Claude Opus 5 and Fable 5 were used in the preparation of this manuscript including checking symbolic derivations in Sec. \ref{sec:analyticUphill} and Appendix \ref{sec:appendixa}, typesetting manuscript equations, preparing input scripts for LAMMPS, and for generating plotting code. 

%%%%%%%%%%%%%%%%%%%%%%%%%%%%%%%%%%%%%%%%%%%%%%%%%%
\section*{Data Availability}

No new data were generated or analyzed in support of this research.



%\bibliographystyle{mnras}

\begin{thebibliography}{}
\makeatletter
\relax
\def\mn@urlcharsother{\let\do\@makeother \do\$\do\&\do\#\do\^\do\_\do\%\do\~}
\def\mn@doi{\begingroup\mn@urlcharsother \@ifnextchar [ {\mn@doi@} {\mn@doi@[]}}
\def\mn@doi@[#1]#2{\def\@tempa{#1}\ifx\@tempa\@empty \href {http://dx.doi.org/#2} {doi:#2}\else \href {http://dx.doi.org/#2} {#1}\fi \endgroup}
\def\mn@eprint#1#2{\mn@eprint@#1:#2::\@nil}
\def\mn@eprint@arXiv#1{\href {http://arxiv.org/abs/#1} {{\tt arXiv:#1}}}
\def\mn@eprint@dblp#1{\href {http://dblp.uni-trier.de/rec/bibtex/#1.xml} {dblp:#1}}
\def\mn@eprint@#1:#2:#3:#4\@nil{\def\@tempa {#1}\def\@tempb {#2}\def\@tempc {#3}\ifx \@tempc \@empty \let \@tempc \@tempb \let \@tempb \@tempa \fi \ifx \@tempb \@empty \def\@tempb {arXiv}\fi \@ifundefined {mn@eprint@\@tempb}{\@tempb:\@tempc}{\expandafter \expandafter \csname mn@eprint@\@tempb\endcsname \expandafter{\@tempc}}}

\bibitem[\protect\citeauthoryear{Andersson \& Ågren}{Andersson \& Ågren}{1992}]{AnderssonAgren1992}
Andersson J.,  Ågren J.,  1992, \mn@doi [Journal of Applied Physics] {10.1063/1.351745}, 72, 1350

\bibitem[\protect\citeauthoryear{{Bildsten} \& {Hall}}{{Bildsten} \& {Hall}}{2001}]{Bildsten2001}
{Bildsten} L.,  {Hall} D.~M.,  2001, \mn@doi [\apjl] {10.1086/319169}, \href {https://ui.adsabs.harvard.edu/abs/2001ApJ...549L.219B} {549, L219}

\bibitem[\protect\citeauthoryear{{Blouin}, {Daligault}  \& {Saumon}}{{Blouin} et~al.}{2021}]{Blouin2021}
{Blouin} S.,  {Daligault} J.,   {Saumon} D.,  2021, \mn@doi [\apjl] {10.3847/2041-8213/abf14b}, \href {https://ui.adsabs.harvard.edu/abs/2021ApJ...911L...5B} {911, L5}

\bibitem[\protect\citeauthoryear{{Burgers}}{{Burgers}}{1969}]{Burgers1969}
{Burgers} J.~M.,  1969, {Flow Equations for Composite Gases}

\bibitem[\protect\citeauthoryear{Bédard, Blouin  \& Cheng}{Bédard et~al.}{2024}]{Bedard2024}
Bédard A.,  Blouin S.,   Cheng S.,  2024, \mn@doi [Nature] {10.1038/s41586-024-07102-y}, 627, 286

\bibitem[\protect\citeauthoryear{{Caplan}}{{Caplan}}{2023}]{Caplan2023RNAAS}
{Caplan} M.~E.,  2023, \mn@doi [Research Notes of the American Astronomical Society] {10.3847/2515-5172/accd6e}, \href {https://ui.adsabs.harvard.edu/abs/2023RNAAS...7...76C} {7, 76}

\bibitem[\protect\citeauthoryear{Caplan \& Freeman}{Caplan \& Freeman}{2021}]{caplan2021mnras}
Caplan M.,  Freeman I.,  2021, \mnras, 505, 45

\bibitem[\protect\citeauthoryear{{Caplan}, {Bauer}  \& {Freeman}}{{Caplan} et~al.}{2022}]{CBF}
{Caplan} M.~E.,  {Bauer} E.~B.,   {Freeman} I.~F.,  2022, \mn@doi [\mnras] {10.1093/mnrasl/slac032}, \href {https://ui.adsabs.harvard.edu/abs/2022MNRAS.513L..52C} {513, L52}

\bibitem[\protect\citeauthoryear{{Chapman} \& {Cowling}}{{Chapman} \& {Cowling}}{1970}]{Chapman1970}
{Chapman} S.,  {Cowling} T.~G.,  1970, {The Mathematical Theory of Non-uniform Gases}

\bibitem[\protect\citeauthoryear{Darken}{Darken}{1942}]{darken1942diffusion}
Darken L.~S.,  1942, Transactions of the American Institute of Mining and Metallurgical Engineers, 150, 157

\bibitem[\protect\citeauthoryear{Darken}{Darken}{1948}]{darken1948diffusion}
Darken L.~S.,  1948, Transactions of the AIME, 175, 184

\bibitem[\protect\citeauthoryear{Darken}{Darken}{1949}]{darken1949diffusion}
Darken L.~S.,  1949, Trans. AIME., 180, 430

\bibitem[\protect\citeauthoryear{Duncan \& Toor}{Duncan \& Toor}{1962}]{duncan1962}
Duncan J.~B.,  Toor H.~L.,  1962, \mn@doi [AIChE Journal] {https://doi.org/10.1002/aic.690080112}, 8, 38

\bibitem[\protect\citeauthoryear{Horowitz, Berry  \& Brown}{Horowitz et~al.}{2007}]{Horowitz2007}
Horowitz C.~J.,  Berry D.~K.,   Brown E.~F.,  2007, \mn@doi [Phys. Rev. E] {10.1103/PhysRevE.75.066101}, 75, 066101

\bibitem[\protect\citeauthoryear{Krishna}{Krishna}{2019}]{Krishna2019}
Krishna R.,  2019, \mn@doi [Chemical Engineering Science] {https://doi.org/10.1016/j.ces.2018.10.032}, 195, 851

\bibitem[\protect\citeauthoryear{Liu, Vlugt  \& Bardow}{Liu et~al.}{2011}]{liu2011}
Liu X.,  Vlugt T.~J.,   Bardow A.,  2011, \mn@doi [Industrial & Engineering Chemistry Research] {10.1021/ie201008a}, 50, 10350

\bibitem[\protect\citeauthoryear{Maxwell}{Maxwell}{1867}]{Maxwell1867}
Maxwell J.~C.,  1867, \mn@doi [Philosophical Transactions of the Royal Society of London] {10.1098/rstl.1867.0004}, pp 49--88

\bibitem[\protect\citeauthoryear{{Mckinven}, {Cumming}, {Medin}  \& {Schatz}}{{Mckinven} et~al.}{2016}]{Mckinven2016}
{Mckinven} R.,  {Cumming} A.,  {Medin} Z.,   {Schatz} H.,  2016, \mn@doi [\apj] {10.3847/0004-637X/823/2/117}, \href {https://ui.adsabs.harvard.edu/abs/2016ApJ...823..117M} {823, 117}

\bibitem[\protect\citeauthoryear{{Medin} \& {Cumming}}{{Medin} \& {Cumming}}{2010}]{Medin2010}
{Medin} Z.,  {Cumming} A.,  2010, \mn@doi [\pre] {10.1103/PhysRevE.81.036107}, \href {https://ui.adsabs.harvard.edu/abs/2010PhRvE..81c6107M} {81, 036107}

\bibitem[\protect\citeauthoryear{{Medin} \& {Cumming}}{{Medin} \& {Cumming}}{2011}]{Medin2011}
{Medin} Z.,  {Cumming} A.,  2011, \mn@doi [\apj] {10.1088/0004-637X/730/2/97}, \href {https://ui.adsabs.harvard.edu/abs/2011ApJ...730...97M} {730, 97}

\bibitem[\protect\citeauthoryear{Stefan}{Stefan}{1871}]{Stefan1871}
Stefan J.,  1871, Sitzungsberichte der Kaiserlichen Akademie der Wissenschaften Wien, Mathematisch-Naturwissenschaftliche Classe, Abteilung II, 63, 63

\bibitem[\protect\citeauthoryear{Taylor \& Krishna}{Taylor \& Krishna}{1993}]{taylor1993multicomponent}
Taylor R.,  Krishna R.,  1993, Multicomponent Mass Transfer.
John Wiley \& Sons, New York

\bibitem[\protect\citeauthoryear{{White}, {Ticknor}, {Meyer}, {Kress}  \& {Collins}}{{White} et~al.}{2019}]{White2019}
{White} A.~J.,  {Ticknor} C.,  {Meyer} E.~R.,  {Kress} J.~D.,   {Collins} L.~A.,  2019, \mn@doi [\pre] {10.1103/PhysRevE.100.033213}, \href {https://scixplorer.org/abs/2019PhRvE.100c3213W} {100, 033213}

\makeatother
\end{thebibliography}

\providecommand{\noopsort}[1]{}\providecommand{\singleletter}[1]{#1}%

%\begin{thebibliography}{99}
%\end{thebibliography}

%%%%%%%%%%%%%%%%%%%%%%%%%%%%%%%%%%%%%%%%%%%%%%%%%%

%%%%%%%%%%%%%%%%% APPENDICES %%%%%%%%%%%%%%%%%%%%%

\appendix

\section{Mixtures with N>3}\label{sec:appendixa}

\begin{figure*}
\centering
\includegraphics[width=\textwidth]{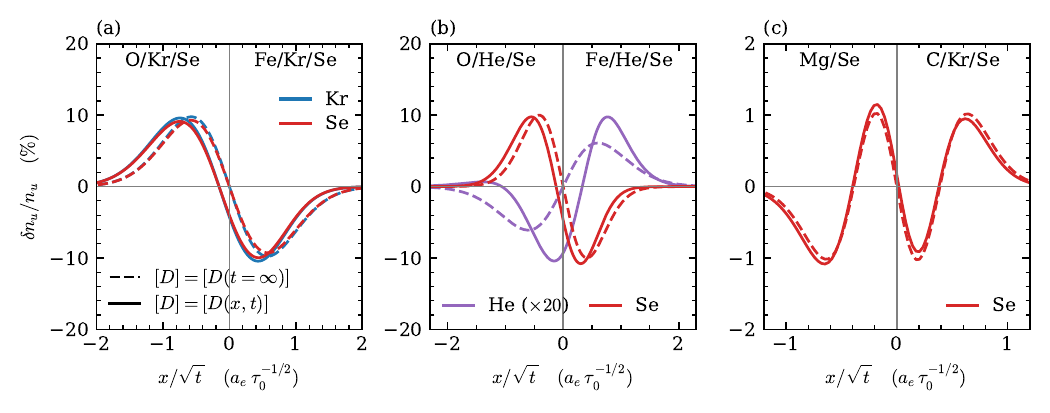}
\caption{Uphill pulses in three mixtures of $N=4$, all computed in the same manner as Fig.~\ref{fig:infmedium} with $[D]=[D(t=\infty)]$ (dashed) and $[D]=[D(z,t)]$ (solid). (a) O/Kr/Se$\,|\,$Fe/Kr/Se: two uphill species of similar charge. (b) O/He/Se$\,|\,$Fe/He/Se: counter-directed uphill diffusion, with He magnified $\times20$. (c) Mg/Se$\,|\,$C/Kr/Se: a double-lobed Se profile. Compositions are given in the text.}
\label{fig:Nspecies}
\end{figure*}

In Sec. \ref{sec:theory}, we present the matrix form of $D_{ij}$ for the special case of $N=3$ (Eq.~\ref{eq:Dij}) but our derivation up until that point was completely general and an expression for the Fick matrix for a mixture of an arbitrary number of components is easily found with the generalized Darken relation to be

\begin{equation}
\begin{split}
\tilde D_{ij}=D_i\,\delta_{ij}-Y_iD_j
-Y_i\left(Z_iD_i-\langle ZD\rangle\right)Z_j^{-1}, \\ \langle ZD\rangle\equiv\sum_kY_kZ_kD_k ,
\label{eq:fickfullB}
\end{split}
\end{equation}

\noindent For the sake of analytic estimates, we call species that are not uniform across the interface the `drivers' while those that are initially uniform across the interface a `driven species' (the uphill species). Requiring that species $N$ be a driver, we can eliminate it by recalling there are only $N-1$ independent concentrations (recall $N=3$ where we used $Y_3 = 1 - Y_1 - Y_2$), so using the general form of $Y_N=1-\sum_{j<N}Y_j$ we obtain the matrix 

\begin{equation}
\begin{aligned}
    \frac{D_{ij}}{D_0} = & Z_i^{-0.6}\delta_{ij}-Y_i\left(Z_j^{-0.6}-Z_N^{-0.6}\right) \\ &-Y_i\Big(Z_i^{0.4}-\sum_{k=1}^{N}Y_kZ_k^{0.4}\Big)\left(\frac{1}{Z_j}-\frac{1}{Z_N}\right)
\end{aligned}\; 
 i,j\in\{1,\dots,N-1\}
\end{equation}

\noindent with $N-1$ eigenvalues $\lambda$. The linear approximations in Sec. \ref{sec:theory} again generalizes and we can define an uphill number for every driven species. Again using the mean mixture approximation (\textit{i.e.} a constant $[D]$ for fully mixed $[D(t=\infty)]$), we will obtain a result analogous to the $N=3$ case, whereby each driven species couples to all drivers on its row of the matrix. Each uphill species has a profile, analogous to Fig.~\ref{fig:infmedium} that is a sum of $N-1$ error functions on the eigenvalues of the matrix, with each eigenvalue contributing a term. 
The driven species couples directly to the gradients only through the off diagonal elements on its own row, \textit{i.e.}, $D_{ij}$ with $i \neq j$. 

For the case of $N=4$, the $3\times3$ matrix contributes two off-diagonal elements to each uphill species. Define $F(\lambda) = \operatorname{erf}(z/2\sqrt{\lambda t})$ so that 

\begin{equation}
\begin{aligned}
& F[\lambda_1,\lambda_2]
=\frac{F(\lambda_1)-F(\lambda_2)}{\lambda_1-\lambda_2},
\\
& F[\lambda_1,\lambda_2,\lambda_3]
=\frac{F[\lambda_1,\lambda_2]-F[\lambda_2,\lambda_3]}
{\lambda_1-\lambda_3}.
\end{aligned}
\end{equation}

\noindent so the pulse shape of an uphill species $i$ becomes 

\begin{equation}
\begin{aligned}
    \delta Y_i(z,t) = -\frac{1}{2}\biggl[
    &\Big(\sum_j D_{ij}\Delta Y_j\Big)
    F[\lambda_1,\lambda_2]
    \\
    &+\big[(D-\lambda_1)(D-\lambda_2)\,\Delta Y\big]_i
    F[\lambda_1,\lambda_2,\lambda_3]
    \biggr].
\end{aligned}
\end{equation}

\noindent where the peak overshoot is again
$0.121\,\Pi_i$ at $z=\pm\sqrt{2\bar\lambda t}$ so long as the eigenvalues are close and the first term dominates, with
\begin{equation}
\Pi_i\equiv\frac{\big|\sum_jD_{ij}\,\Delta Y_j\big|}{Y_i\,\bar\lambda},
\qquad
\bar\lambda=\frac{\mathrm{Tr}[D]}{N-1},
\label{eq:PiN}
\end{equation}
which again reduces to Eq.~(\ref{eq:Pi}) for $N=3$. As before, $\Delta Y_j$ is the difference in concentration on the left and right and so is 0 for our driven species. 

Observe that going from $N=3$ to $N=4$ introduces many new possible pulse shapes that are amusing and sometimes counterintuitive. A few illustrative mixtures are described below, and their pulse shapes computed numerically from the Fick matrix (but not MD) are shown in Fig.~\ref{fig:Nspecies}.%We show a few below, mostly to illustrate how a third driving species modifies the pulse shapes of the one uphill species in amusing and sometimes counterintuitive ways. 

To begin, consider the simplest extension of our fiducial mixture in $N=4$: a second uphill species with two drivers can be obtained by taking our original O/Fe/Se mixture and replacing half of the Se with another species of similar high charge like Kr. Both species would be expected to produce an excess in the O of similar amplitude (about 10\%), shown in Fig.~\ref{fig:Nspecies}(a).

However, choosing a very low $Z$ uphill species, significantly less than O, could produce counter-directed uphill diffusion, where one species produces an excess on the left while the other produces an excess on the right. Some quick numerical spot checks show that this is possible by replacing 0.05 of the Se charge fraction with He in our fiducial O/Fe/Se mixture. Formally, this He/O/Fe/Se mixture with $\mathbf{Z}=(2, 8, 26, 34)$ has $Y^{L} = (0.05, 0.7, 0, 0.25)$ and $Y^{R} = (0.05, 0, 0.7, 0.25)$. Estimates using the equations above suggest that the He excess on the Fe side is at the sub-percent level while the Se remains at the 10\% level, so the uphill He signal is not easily resolvable with MD, and it probably also has larger errors to due to the CBF law being taken to such large charge ratios; see Fig.~\ref{fig:Nspecies}(b). From this first check it seems like counter-uphill diffusion is suppressed in Yukawa mixtures, but it suggests that one might be able to contrive a mixture with detectable counter-uphill diffusion if simulating with very large charge ratios. We conjecture that the added parameters for $N=5$ and by using charges much larger than Se that the effect can be tuned to strengthen this signal but probably not by much.

As a final example in $N=4$ we consider the case of three drivers and one uphill species. The additional error function also allows the uphill profile to change signs on both sides of the interface, producing a seemingly paradoxical `doubly uphill' signal which has excesses on both sides of the interface relative to its initial composition; this paradox is resolved by recognizing that it is not due to the inventory growing on both sides, but rather, the differences in error functions producing enhanced and depleted pulses on both sides of the interface. One such example is shown in Fig.~\ref{fig:Nspecies}(c) using an uphill Se with $Y_{\rm Se} = 0.3$ on both sides, with the left driver being Mg ($Y_{\rm Mg} = 0.7$) and the right drivers being C ($Y_{\rm C} = 0.386$) and Kr ($Y_{\rm Kr} = 0.314)$.

%%%%%%%%%%%%%%%%%%%%%%%%%%%%%%%%%%%%%%%%%%%%%%%%%%

% Don't change these lines
\bsp	% typesetting comment
\label{lastpage}
\end{document}